\documentclass[12pt]{article}
\usepackage[utf8]{inputenc} 
\usepackage{graphicx}
\usepackage{times}
\usepackage{sectsty}
\usepackage{titlesec}
\usepackage{float}
\usepackage{hyperref}
\usepackage{caption}
\usepackage{listings}
\usepackage{wrapfig}
\usepackage{subfig}
\usepackage{cite}
\usepackage{xcolor, soul}
\sethlcolor{green}

\usepackage{array,ragged2e} 
\usepackage{booktabs}
\usepackage{indentfirst}
\allsectionsfont{\normalsize\bfseries}
\usepackage[letterpaper, margin=1in]{geometry}

\titlespacing\section{0pt}{12pt plus 2pt minus 0pt}{5pt plus 2pt minus 2pt} 
\titlespacing\subsection{0pt}{12pt plus 2pt minus 0pt}{5pt plus 2pt minus 2pt}

\usepackage{titling}
\title{\large \textbf{Perceived Importance of Cognitive Skills Among Computing Students in the Era of AI}}
\date{}
\author{Neha Rani, Erta Cenko, and Laura Melissa Cruz Castro\\
\normalsize University of Florida, Gainesville, FL, USA}

\begin{document}
\maketitle
\vspace{-20pt}
\section*{Abstract}
\noindent
The availability and increasing integration of generative AI tools have transformed computing education. While AI in education presents opportunities, it also raises new concerns about how these powerful know-it-all AI tools, which are becoming widespread, impact cognitive skill development among students. Cognitive skills are essential for academic success and professional competence. It relates to the ability to understand, analyze, evaluate, synthesize information and more. The extensive use of these AI tools can aid in cognitive offloading, freeing up cognitive resources to be used in other tasks and activities. However, cognitive offloading may inadvertently lead to diminishing cognitive involvement in learning and related activities when using AI tools. Understanding cognitive skills' impact in the era of AI is essential to align curricular design with evolving workforce demands and changing work environment and processes. To address this concern and to develop an understanding of how the importance of cognitive skills changes with increasing integration of AI, we conducted a researcher-monitored and regulated quantitative survey of undergraduate computing students. We examined students' perceptions of cognitive skills across three temporal frames: prior to widespread AI adoption (past), current informal and formal use of AI in learning contexts (present), and future with even more AI integration in professional environments (future). In the study, students rated the importance of 11 cognitive skills, including critical thinking, problem solving, decision making, attention control, working memory, and cognitive flexibility. Our analysis reveals that students expect all 11 cognitive skills to be of diminishing importance in the future, when AI use and integration increases. Although literature and educators often emphasize the enduring importance of cognitive skills in computing education, our investigation suggests that students may view these skills as less necessary in an AI-rich environment. Survey data indicate that computing students anticipate a significantly diminished role for critical cognitive skills, such as problem solving, abstract reasoning, and cognitive flexibility, in the future. This perception risks under-preparing students for professional contexts where cognitive skills remain crucial. The importance of these skills is likely to continue, even more so with the AI presence, as computing learners and professionals need to be able to assess AI outcomes and use them appropriately. Our findings highlight the need for educational interventions that explicitly reinforce cognitive skill development within learning environments that are now often relying on AI.

\section{Introduction}
\noindent
The landscape of engineering and computing is rapidly evolving, while witnessing the rapid integration of AI to streamline various processes in the workplace and education \cite{abrahao2025software, liu2025application}. We live in uncertain times, where the evolution of the workspace integrated with AI is still emerging and taking its form. This creates a highly uncertain environment for preparing computing professionals who are soon to enter the workforce, such as university students enrolled in engineering and computing programs. Moreover, even the learning space is observing a surge in AI usage \cite{liu2025application}. While there are some formal uses, much of the use remains informal and therefore unregulated, further exacerbating the challenge of ensuring that future computing professionals develop the right set of cognitive skills. Multiple national surveys have emphasized the importance of cognitive skills in engineering education \cite{director2004engineer}. Various cognitive skills, such as problem solving, reasoning, and critical thinking, have long been considered critical in higher education \cite{sankar2008use}. The future of the nation depends on the technological progress and knowledge advancements made in the field of engineering \cite{al2019modeling}. This calls for attention and thorough investigation to understand what skill sets are essential to be developed. This will help decide the area of concern that engineering education should shift its focus to. To better understand the shifting cognitive skill needs and perceptions, we conduct a study with students enrolled in the engineering program at a large US public university.
\bigbreak
\noindent
Developing cognitive skills among engineering students is a crucial pedagogical objective. Cognitive skills equip them with the skillset to interpret complex information, evaluate trade-offs, anticipate unintended consequences, and design solutions that are safe, ethical, and sustainable. 
Cognitive skills entail the mental abilities required to perform tasks ranging from simple to complex, including information handling, judgment, awareness, memory, and reasoning \cite{sciencedirectCognitiveSkill}.
Cognitive skills such as critical thinking, problem solving, decision making, abstract reasoning, creativity, metacognition, cognitive flexibility, information literacy, system thinking, and verbal reasoning are crucial for engineers \cite{hames2015study, smith2020cognitive}. Engineers are often involved in creating solutions to complex problems, which can involve analyzing the problem, finding a solution through previous knowledge, past experience, reflective thinking, and applying methods.
A more formal definition of critical thinking is “the intellectually disciplined process of
actively and skillfully conceptualizing, applying, analyzing, synthesizing, and/or evaluating information gathered from, or generated by, observation, experience, reflection, reasoning, or communication, as a guide to belief and action” \cite{criticalthinkingDefiningCritical}. Critical thinking is a skill developed over time and refined daily by practice. It allows an individual to solve complex problems \cite{snyder2008teaching}.
Critical thinking has been emphasized as a critical skill during education by various education boards and accreditation boards, such as ABET\cite{ahern2019literature, shepell2018navigating}. 
It is a vital skill recognized both in industry and academia for engineering graduates \cite{ahern2012critical, ahern2019literature}. These skills will help them perform better in a job, solve complex problems, and transfer their skills to apply their knowledge to different scenarios. With the use of AI, the processes involved in solving problems are now changing, often by delegating tasks to AI. Thus, the use of AI will impact critical thinking skills. 
\bigbreak
\noindent
Engineers need working memory and attention control, which are parts of cognitive skills, to focus on solving the problem at hand. Working memory allows individuals to hold, store, and manipulate information for a short period of time to aid in problem-solving. Attention control is the ability that allows individuals to maintain focus on relevant stimuli and shift focus as required. Often, working with AI means changing focus from the problem to how to interact with AI, such as concerns for creating the right prompts. These interactions with AI during problem-solving may negatively impact attention control and available working memory. Abstract thinking is another cognitive skill that may be negatively impacted by the unregulated use of AI. Abstract reasoning is the ability to analyze problems of situation, find emerging patterns, and find solutions on a conceptual level.
Systems thinking is another skill that indicates how well an individual understands the system's internal function and how its components work together. This is essential for engineers as they often work with multiple complex systems and are part of the design and development of systems. The quality of thinking often determines how they operate the system and what design solutions they develop \cite{ahern2019literature}. The prevalence of AI and the use of more and more black-box systems during the educational phase may undermine their system thinking skills. Metacognition is another cognitive skill that is concerned with awareness and regulation of one's own thoughts. With the ease of availability of AI, engineering students are constantly making choices between an AI and a non-AI solution approach. Therefore, the presence of AI demands more focus on metacognition \cite{gerlich2025ai}.
\bigbreak
\noindent
Further, creativity as a cognitive skill is also likely to be impacted due to AI. Some prior work mentions that the use of AI speeds up creative ideas and work by simply collaborating with individuals \cite{abrahao2025software}. However, some researchers indicate that the increase in the use of AI could mean more cognitive offloading and therefore less indulgence in the tasks and reflective thinking, and therefore reduce creativity \cite{gerlich2025ai}. Cognitive flexibility is also suggested to be negatively impacted by AI \cite{gerlich2025ai}. Cognitive flexibility is a cognitive skill that determines one's ability to switch between tasks \cite{ionescu2012exploring}. Omnipresent AI also means humans are constantly working with AI and often collaborating with AI, which also means decision-making is done collaboratively. Further, overreliance on AI for decision making is also observed \cite{gerlich2025ai}. This is likely to impact the decision-making skills of individuals.
\bigbreak
\noindent
Overall, it seems like AI is going to have a potential impact on multiple cognitive skills and their development. Given the pressing need to prepare a computing workforce for the changing workplace scenario, which is moving towards being an AI-rich environment. To address this pressing need and contribute towards understanding how the needs of cognitive skills development are being shaped due to the presence of AI, we conducted a study with students enrolled in an engineering program at the undergraduate level at a large public US university with an intensive research program and AI endeavors. This investigation is primarily based on student perception. Students are the key stakeholders, and understanding their perception is crucial. To ensure gathering perception from students with ample AI exposure, we recruited participants who were sophomores or higher level in their degree program.


\section{Background and Related Work}
\noindent
Exceedingly fast development and adoption of AI tools into educational environments have left old pedagogical practitioners questioning if they are still teaching useful skills to students, especially within engineering disciplines \cite{ozer2024adapting}. Because of these tools, a shift is needed in engineering education practices with emphasis on cognitive skills required in the current world context. 
Large Language Models and Generative AI continue to be at students' disposal, allowing them to delegate important parts of their own exercises, such as writing code, concept understanding, and problem solving \cite{wermelinger2023using}. As use of these tools continues to become widespread across computing education and becomes more capable, it becomes important to understand the impact on student learning and cognitive skill development, as they are the computing professionals soon to join the workforce \cite{zastudil2023generative}. 
\bigbreak
\noindent
Cognitive skills such as critical thinking, problem-solving, and decision-making have long been the goals of engineering education and of skill development in computing education. 
The presence and usage of AI in educational spaces and workspaces are likely to impact these cognitive skills, either negatively or positively. Currently, we have a limited understanding of AI's impact on cognitive skills, as this is still an emerging area of research with limited prior work and empirical evidence. Further, cognitive skills are shaped or impacted over a long period of time. So a longitudinal investigation is needed to make a concrete conclusion. Various authors have shared their position and possible outcomes of rapid integration of AI while little has been found through empirical investigations. Integrating AI can streamline processes in the software development workspace, speed up work, and support creativity while reducing friction \cite{abrahao2025software}. It may decrease developers' understanding of the codebase due to reduced codebase interaction and more AI interaction \cite{abrahao2025software}. AI integration can impact critical thinking through cognitive offloading \cite{gerlich2025ai}. Cognitive offloading occurs when individuals rely on external aids to handle mental tasks, which reduces their involvement in deep, reflective thinking \cite{gerlich2025ai}.
 The adverse impact of AI usage on critical thinking skills has been observed multiple times \cite{gerlich2025ai}. Some of the prior work also points towards positive outcomes of the inclusion of AI. They mention AI could free up individuals from redundant tasks by outsourcing them to AI, while they themselves can focus on in-depth algorithmic understanding and decision-making \cite{becker2023programming}. Another work suggests that the use of AI will likely promote and support creativity in open education \cite{mills2023we}.


\subsection{Research Questions}
\noindent
With the introduction and widespread popularity, and the push from the government and industry to build a more AI-rich environment, as well as the need to train upcoming professionals to better prepare them for the AI era that is continuously evolving, this study aims to understand how engineering students perceive the importance of cognitive skills. More specifically, we want to contribute towards the understanding of how the importance of these cognitive skills is being shaped with the presence of AI, and as time evolves, with more and more inclusion of AI. Through our investigation, we aim to answer the following research questions.
\bigbreak
\noindent
\begin{itemize}
\item \textit{\textbf{RQ1: What is the computing students' perceived importance of different cognitive skills for computing professionals?}}

\item \textit{\textbf{RQ2: Does the perception of importance of different cognitive skills for computing professionals change if we compare past, present, and future times?}}
\bigbreak
\end{itemize}

\section{Methodology}
It was a within-subjects study design. It was a survey-based controlled study involving one-to-one sessions with the researcher.
The study was approved by the university's IRB. The study was conducted by two researchers. The study involved scheduling a time slot with the participants. The researcher then guided the participant over the teleconferencing portal to complete the study. The participants were university students enrolled in a computer science course. Participants were compensated with 0.5 course credits for their participation. The study took approximately 30 minutes to complete. This was a survey-based study conducted in an experimental setup under the supervision of a researcher.
\subsection{Procedure}
\begin{figure}
    \centering
    \includegraphics[width=0.85\linewidth]{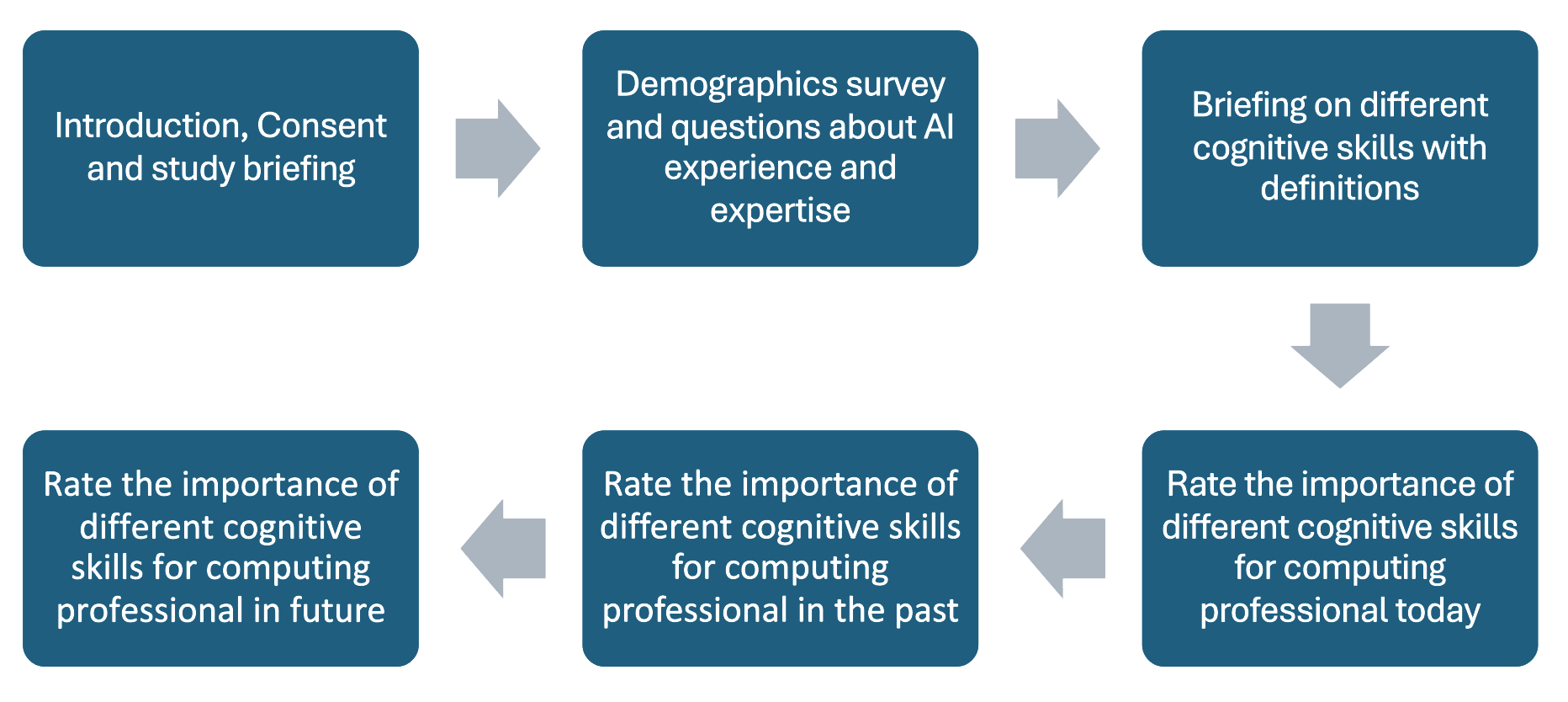}
    \caption{Study procedure}
    \label{Study procedure}
\end{figure}

\noindent
The study began with an introduction and a briefing of the students about the study. Followed by which they provide the consent form. Upon signing the consent form, preliminary survey was sent for completion. This survey focuses on the demographics and their prior experience. To help understand the upcoming cognitive skills questions, students were first provided with definitions of the cognitive skills they were going to be asked about. Then they were shared the survey link for response regarding their perceived importance of the cognitive skills for computing professionals in the present.
Then they were asked to imagine it was 2015, and LLMs such as ChatGPT we see today did not exist then. Then they were asked to answer the same set of questions about the importance of cognitive skills for computing professionals, but in the past, the pre-AI era.
In the next step, participants were asked to imagine a future where AI is more integrated. Then they were asked to answer the same set of questions about the importance of cognitive skills for computing professionals, but in a future, more AI-integrated environment. 
The study procedure is delineated in figure \ref{Study procedure}.
The entire study was conducted during a scheduled session, during which the researcher guided the participant through the study steps.
The cognitive skills that were assessed in the survey are listed and defined below:
\begin{itemize}

 \item Critical Thinking: The ability to evaluate information objectively and make reasoned judgments. 

 \item Problem Solving: The ability to find solutions to difficult or complex issues.

 \item Decision-Making: The process of choosing the best course of action among alternatives.

 \item Attention Control: The capacity to maintain focus on relevant stimuli and shift focus as needed. 

 \item Working Memory: The ability to hold and manipulate information over short periods. 

 \item Cognitive Flexibility: The ability to switch between different tasks or mental frameworks. 

 \item Abstract Reasoning: The ability to analyze information, detect patterns, and solve problems on a complex, conceptual level. 

 \item Metacognition: Awareness and regulation of one’s own thought processes. 

 \item Information Literacy: The ability to identify, evaluate, and use information effectively. 

 \item Creativity: The ability to produce novel and useful ideas. 

 \item {Systems Thinking: Understanding how parts of a system interact and influence one another.}


\end{itemize}

\subsection{Participants}
\noindent
There were a total of 17 participants, out of which 10 were male and 7 were female. Of 17, 7 were White, 9 were asian, and 1 was african american. The participant pool was a mix of students from higher levels in their degree program. We made sure to recruit students with some experience in their degree program so that they have some level of exposure to AI and the engineering curriculum. There were 6 sophomores, 9 juniors, and 2 seniors. All these students were enrolled in a computing degree program, such as Computer Engineering or Computer Science.

\subsection{Analysis}
\noindent
To understand computing students' perception of the current importance of different cognitive skills for computing professionals, a descriptive analysis of the data was run.
A sample cognitive skill importance question was "Rate critical thinking in terms of its current importance for computing professions." Rate on a scale of 1 to 10, where 1 is not important at all, and 10 is very important.
We first ran the Shapiro-Wilk normality test on the data.
To understand differences in computing students' perceptions of the importance of cognitive skills, we ran a one-way repeated-measures ANOVA or the nonparametric alternative, the Friedman test.
The one-way repeated-measures ANOVA or the nonparametric alternative, the Friedman test, was run individually on all 11 cognitive skills importance responses.
Further, we ran an Exploratory factor analysis to examine relationships among the different cognitive skills.

\section{Results}
\begin{figure}
    \centering
    \includegraphics[width=0.5\linewidth]{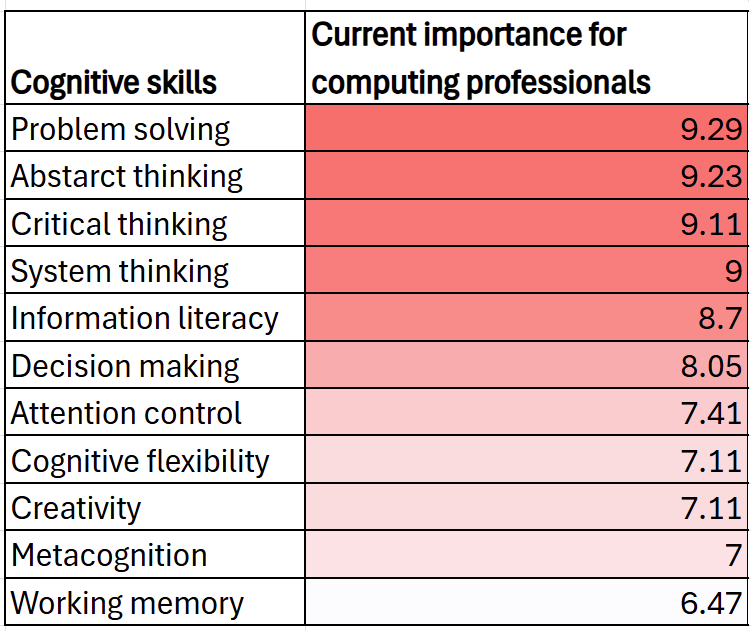}
    \caption{Cognitive skills importance ratings with heatmap color coding. A higher color gradient indicates higher ratings. Ratings are on a scale of 1 to 10, where 1 is not important at all, and 10 is very important.}
    \label{cognitive skills current}
\end{figure}

\noindent
Problem solving was rated as the most important skill with a mean rating of 9.29 on a scale of 1 to 10 (SD=1.26). 
Followed by abstract reasoning as another important skill, with a mean rating of 9.23 on a scale of 1 to 10 (SD=0.83).
Then, critical thinking was rated as another important skill, with a mean rating of 9.11 on a scale of 1 to 10 (SD=1.11).
While working memory was rated as the least important cognitive skill, with a mean rating of 6.47 on a scale of 1 to 10 (SD=1.58).
Metacognition was another cognitive skill rated as less important, with a mean rating of 7.00 on a scale of 1 to 10 (SD=1.80).
Creativity is also rated as less important with a mean of 7.11 (SD=2.34).
Overall, all 11 cognitive skills were rated above the median.
All the scores are shown in Figure \ref{cognitive skills current}.
                            
\bigbreak
\noindent
Results of the one-way repeated measures ANOVA show that the importance of problem-solving changes significantly across time, i.e., the past, present, and future (p$<$.001). Results show that problem-solving skill is very important in the past (mean=9.52, SD=0.71) and current (mean=9.29, SD=1.26), but will become much less important in the future (mean=6.52, SD=2.47).
A one-way repeated-measures ANOVA on working memory also shows a significant difference in its importance for computing professionals across time (past, present, and future) (p$<$.001). Working memory is considered important in the past (mean=8.00, SD=1.17) and less important in the present (mean=6.47, SD=1.58), and will become significantly less important in the future (mean=5.47, SD=2.42).
Furthermore, even the importance of cognitive flexibility significantly declines over the period of time (p=.01). The Importance of cognitive flexibility in the past (mean=7.81, SD=1.84) is higher and decreases in the present (mean=7.11, SD=1.53) and future (mean=6.35, SD=1.93).
Ratings of the importance of abstract thinking also reduced significantly over time (p$<$.001). Abstract thinking is considered important in the past (mean=9.11, SD=1.36) and a little more important in the present (mean=9.23, SD=.83), and will become significantly less important in the future (mean=6.47, SD=2.96).
Results of the one-way repeated measures ANOVA show that the importance of metacognition changes significantly across time, i.e., the past, present, and future (p=.04). Results show that metacognition is very important in the past (mean=7.58, SD=2.15) and then importance decreases for current time (mean=7.00, SD=1.80), but will become much more important in the future (mean=8.05, SD=1.74).
One-way repeated-measures ANOVA also shows that the importance of information literacy decreases significantly over time (p=.03). The importance of information literacy decreased from the past (mean=8.82, SD=1.55) to the present (mean=8.70, SD=1.64), and decreased further in the future (mean=7.29, SD=2.64).
Results of the one-way repeated measures ANOVA show that the importance of system thinking changes significantly across time (p=.01).
Results show that the importance of system thinking remains similar in the past (mean=9.05, SD=1.02) and present (mean=9.00, SD=1.27) but diminishes significantly in the future (mean=7.76, SD=1.95).
The results os one-way repeated measures ANOVA on the importance of cognitive skills across time is illustrated in Table \ref{tab}

\begin{table}[H]
\centering
\caption{Importance of cognitive skills across past, present, and future. The * shows the p values$<$.05.}
\begin{tabular}{|p{0.25\textwidth}|p{0.15\textwidth}|p{0.15\textwidth}|p{0.15\textwidth}|p{0.1\textwidth}|}
\hline
\multicolumn{1}{|c|}{\textbf{Cognitive skill}} & 
\multicolumn{1}{|p{0.15\textwidth}|}{\textbf{Importance in past}} & 
\multicolumn{1}{|p{0.15\textwidth}|}{\textbf{Importance in present}} & 
\multicolumn{1}{|p{0.15\textwidth}|}{\textbf{Importance in future}} &
\multicolumn{1}{|p{0.1\textwidth}|}{\textbf{p value}}\\
\hline
Problem solving & mean=9.52, SD=0.71 & mean=9.29, SD=1.26 & mean=6.52, SD=2.47 & *p$<$.001\\
\hline
Working memory & mean=8.00, SD=1.17 & mean=6.47, SD=1.58 & mean=5.47, SD=2.42 & *p$<$.001\\
\hline
Cognitive flexibility & mean=7.81, SD=1.84 & mean=7.11, SD=1.53 & mean=6.35, SD=1.93 & *p=.01\\
\hline
Abstract thinking & mean=9.11, SD=1.36 & mean=9.23, SD=.83 & mean=6.47, SD=2.96 & *p$<$.001\\
\hline
Metacognition & mean=7.58, SD=2.15 & mean=7.00, SD=1.80 & mean=8.05, SD=1.74 & *p=.04 \\
\hline
Information literacy & mean=8.82, SD=1.55 & mean=8.70, SD=1.64 & mean=7.29, SD=2.64 & *p=.03 \\
\hline
System thinking & mean=9.05, SD=1.02 & mean=9.00, SD=1.27 & mean=7.76, SD=1.95 & *p=.01 \\
\hline
Critical thinking & mean=9.29, SD=1.35 & mean=9.11, SD=1.11 & mean=8.35, SD=2.20 & p=.07 \\
\hline
Decision making & mean=8.17, SD=1.81 & mean=8.05, SD=1.88 & mean=8.00, SD=1.73 & p=.92 \\
\hline
Attention control & mean=8.05, SD=1.60 & mean=7.41, SD=1.93 & mean=7.17, SD=2.12 & p=.26 \\
\hline
Creativity & mean=8.23, SD=2.27 & mean=7.11, SD=2.34 & mean=8.29, SD=1.68 & p=.10\\
\hline
\end{tabular}
\label{tab}
\end{table}

\section{Discussion}
\noindent
Our analysis shows that all 11 cognitive skills assessed in the study are rated as important in current times, except for working memory, as shown in Figure \ref{cognitive skills current}. While all the skills are rated above median, some skills are rated more important than others. We observe that problem solving, abstract thinking, critical thinking, and systems thinking are rated much higher, i.e., 9 or above on a scale of 1 to 10. 
Interestingly, working memory was rated the lowest yet above median in terms of its importance for computing professionals. This can be somewhat related to cognitive offloading with the use of AI \cite{gerlich2025ai}. The low importance rating of working memory can be attributed to increased use of AI for trivial to complex tasks, making remembering things for a short period less important, as one can quickly ask anything to AI and delegate tasks to it. While this is the perception, it creates a concern for educators and policymakers. While one can ask AI anything they want and delegate a task, still working on a problem requires working memory to be able to triangulate multiple pieces of information, make decisions, and form solutions.
\bigbreak
\noindent
Overall, we observed a diminishing importance of almost all cognitive skills, with a few remaining at a marginally reduced level. We observed that 10 of the 11 cognitive skills assessed became less important. This can be related to the rampant and widespread use of AI and the overreliance that students may have on AI \cite{pitts2025students, pitts2026trust}. Almost all the skills become less important, and many become significantly less important. While this study is only based on computing students' perception, it is still a noteworthy perception of a major stakeholder, as the future computing workforce. This shows an issue we are likely to face with the emergence and widespread integration of AI. 
Reduced importance of system thinking in the future makes sense as AI becomes popular and widespread, computing professionals will already be aware of how the systems function and will not be primarily concerned about it. It can also be seen from another perspective. As AI continues to grow, AI systems will become widely accepted, and computing professionals may no longer be concerned about their functioning and logic. AI will likely become an unquestioned companion, rendering the need for understanding system functioning obsolete. 
A significant decrease in information literacy can be explained by increased reliance on AI systems to provide information as and when it is needed. This also explains the reduction in working memory, as computing professionals will rely on AI to remember and provide information, so humans do not have to worry about memory. A significant reduction in the importance of problem solving skills is potentially due to the use of AI to analyze problems and solve them. Surprisingly, the importance of cognitive flexibility also reduces in the future, which is the ability to switch between different thinking frameworks. It seems like students also expect AI to assist with many tasks, so they do not have to switch between different mental frameworks. Significant reduction in abstract thinking can also be related to the assistance and use of AI in information analysis, pattern detection, and application. Only metacognition was considered to be significantly more important in the future. Possibly with the presence of AI, computing professionals need to become more aware of how they are thinking to keep up with AI and stay ahead of it to remain relevant and employable.

\bigbreak
\noindent
We also observe that the importance of skills such as critical thinking does not change significantly over time. This is reassuring that computing students recognize that critical thinking remains important, and its importance does not diminish even in an AI-rich workspace for computing professionals. It should be noted that while the importance ratings do not change significantly, they do decrease slightly over time. Apart from critical thinking skills, decision making, and attention control importance does not reduce significantly in the future. Decision making will still remain an important skill for computing professionals who work with AI, as they will be making decisions even though AI will assist them in decision making, hence the somewhat reduction in importance. Attention control will continue to be important, with a somewhat reduction in importance, as in an AI-rich environment, computing professionals will have to shift their focus to AI as needed during the problem-solving task. The importance of creativity will increase in the future, although not significantly, possibly due to creativity and creating innovative solutions remaining a human-specific skill, which will be difficult for machines to imitate.
\bigbreak
\noindent
Our investigation will help engineering education instructors and curriculum designers to understand the changes needed in their instruction and assessment accordingly to ensure the holistic development of cognitive skills in the era of AI.

\section{Limitations and Future Work} 
\noindent
One limitation of the study is that it was conducted with computing students who might not be experts in computing. However, to ensure some level of expertise, we recruited students who are at the sophomore level or higher in their degree program to ensure that they have some degree of AI exposure and experience. Additionally, we believe understanding students' perspectives is also important as they are a major stakeholder. While their perception might not impact how the computing industry evolves or how engineering education adapts, it definitely helps researchers, instructors, and industry understand the current mindset of computing students. 
This serves as a direction for future research. In our future work, we hope to conduct an investigation closely with other stakeholders, such as engineering instructors and industry professionals, to closely understand the evolving space of how the importance of cognitive skills is shaped in the evolving AI workspace.
Another limitation of the study is the relatively smaller number of participants despite the survey-based study. Another limitation of the study was the perception report,e and no objective data. We believe it does not impact the quality of results, firstly, as the survey design is coupled with researcher directed and supervised study sessions, ensuring a higher quality of response. Secondly, individual participants were closely explained the different cognitive skills that allowed for the standardization of cognitive skills understanding. This one-to-one session ensures they can ask for clarifications if needed. Thus, further reducing extraneous variables and individual differences in understanding.

\section{Conclusion}
\noindent
Through our investigation, we aim to understand the dynamically changing space of AI integration in computing and its impacts on the importance of cognitive skills and, therefore, the reshaping of the future computing workforce. We made some interesting findings through our researcher-regulated one-to-one survey-based study. Our results are based on empirical evidence from engineering students' perceptions. We found that all 12 cognitive skills assessed in the study are considered important. Some of the most important cognitive skills, as identified in the study, are problem-solving, critical thinking, abstract thinking, and systems thinking. We also found diminishing perception of the importance of almost all cognitive skills in the future with more AI integration. Importance of cognitive skills such as problem solving, working memory, cognitive flexibility, abstract thinking, information literacy, and systems thinking reduces significantly over time. Overall, our investigation provides initial evidence of an emerging problem that engineering education instructors and stakeholders need to address.

\section{Acknowledgments}
We would like to acknowledge Jeevan Ram Munnangi for his contribution to the literature review. We would also like to thank all the students who participated in the study.

\bibliography{ref}
\end{document}